\documentclass[epsf,prb,twocolumn,nofootinbib]{revtex4}

\usepackage[pdftex]{graphicx}
\usepackage{dcolumn}
\usepackage{bm}
\usepackage{epsfig}
\usepackage{latexsym}
\usepackage{amsmath}
\usepackage{amsfonts}
\usepackage{amssymb}
\usepackage{color}
\usepackage{array}
\usepackage{framed}
\usepackage{verbatim}

\usepackage[T1]{fontenc}
\usepackage{ae,aecompl}

\newcommand{\vpi}{{\bm{\pi}}}

\newcommand{\xv}{{\bf x}}

\newcommand{\zh}{{\hat{\bf z}}}

\newcommand{\oh}{{\frac{1}{2}}}

\newcommand{\cH}{{\mathcal H}}

\newcommand{\G}{{\bf G}}
\newcommand{\grad}{{\bm{\nabla}}}

\newcommand{\be}{\begin{equation}}
\newcommand{\ee}{\end{equation}}
\newcommand{\bea}{\begin{eqnarray}}
\newcommand{\eea}{\end{eqnarray}}
\newcommand{\bse}{\begin{subequations}}
\newcommand{\ese}{\end{subequations}}
\def\rf#1{(\ref{#1})}

\begin{document}

\title{Symmetry Enriched Fracton Phases from Supersolid Duality}
\author{Michael Pretko and Leo Radzihovsky}
\affiliation{Department of Physics and Center for Theory of Quantum Matter \\ University of Colorado, Boulder, CO 80309}
\date{\today}

\begin{abstract}
Motivated by the recently established duality between elasticity of crystals and a fracton tensor gauge theory, we combine it with boson-vortex duality, to explicitly account for bosonic statistics of the underlying atoms.  We thereby derive a hybrid vector-tensor gauge dual of a supersolid, which features both crystalline and superfluid order.  The gauge dual describes a fracton state of matter with full dipole mobility endowed by the superfluid order, as governed by ``mutual'' axion electrodynamics between the fracton and vortex sectors of the theory, with an associated generalized Witten effect.  Vortex condensation restores $U(1)$ symmetry, confines dipoles to be subdimensional (recovering the dislocation glide constraint of a commensurate quantum crystal), and drives a phase transition between two distinct fracton phases.  Meanwhile, condensation of elementary fracton dipoles and charges, respectively, provide a gauge dual description of the super-hexatic and ordinary superfluid.  Consistent with conventional wisdom, in the absence of crystalline order, $U(1)$-symmetric phases are prohibited at zero temperature via a mechanism akin to deconfined quantum criticality.
\end{abstract}
\maketitle

\emph{Introduction.}  Many familiar symmetry-breaking phases of matter possess a useful dual description in the language of gauge theory.  Perhaps the most well-known example is boson-vortex duality, which describes two-dimensional superfluids in terms of a Maxwell gauge theory.\cite{stone,dasgupta,fisher} In this mapping, the vortices and Goldstone mode of the superfluid correspond to charges and photon in the Coulomb phase of the gauge theory, respectively.  Meanwhile, a Mott insulating phase of bosons can be described as the Higgs phase in the dual gauge theory.

More recently, it has been realized that two-dimensional elasticity theory has a similar relationship with a rank-two tensor gauge theory hosting fracton excitations.\cite{elasticity} (We refer the reader to a recent review\cite{review} and to selected literature\cite{chamon,haah,fracton1,fracton2,sub,genem,glassy,
spread,mach,holo,deconfined,hanlayer,sagarlayer,field,generic,entanglement,
albert,higgs1,higgs2,fractalsym} for an overview of fractons.)  Phonons map onto the gapless gauge modes, while disclinations and dislocations map onto fracton charges and dipoles, respectively.  The familiar mobility restrictions of lattice defects\cite{emit1,emit2,emit3,emit4} are neatly encoded in higher moment charge conservation laws of the tensor gauge theory.  We note the existence of several related gauge theory dualities concerning elasticity theory.\cite{gromov,pai,beekman1,beekman2}

While fracton-elasticity duality\cite{elasticity} provides a useful description of a Mott-insulating ``commensurate'' crystal (with gapped vacancy/interstitial defects) and offers an insightful embodiment of fracton phenomena, it is not immediately clear how such a formulation can capture a quantum melting transition to non-crystalline phases of the underlying atoms.  Specifically, the tensor-only gauge theory\cite{elasticity} is incomplete as it does not incorporate the quantum statistics of the constituent particles, which is essential in a fluid state.  The tensor gauge theory thus cannot account for the off-diagonal order of the underlying bosonic atoms seen in superfluid and supersolid\cite{Andreev,MosesChen,ketterle,sonss} phases.  In fact, the subdimensional nature of dipoles (absence of dislocation climb\cite{landau,chaikin}) crucially relies on atom number conservation and the associated $U(1)$ symmetry. The deficiency is most striking as one considers a quantum fluid ground state driven by condensation of topological lattice defects (dislocations and/or disclinations). The tensor gauge theory gives no indication why such a fluid generically must exhibit a broken $U(1)$ symmetry associated with superfluid order, seemingly allowing melting into a fully symmetric gapped state.  However, in a continuum (or at incommensurate fillings) such a ``normal'' quantum liquid is forbidden by the Lieb-Schultz-Mattis theorem.\cite{lsm,hastings,oshikawa}  Thus, a complete theory must encode a mechanism tying together the translational and $U(1)$ symmetries precluding their simultaneous preservation in the quantum ground state.

\begin{figure}[t!]
 \centering
 \includegraphics[scale=0.36]{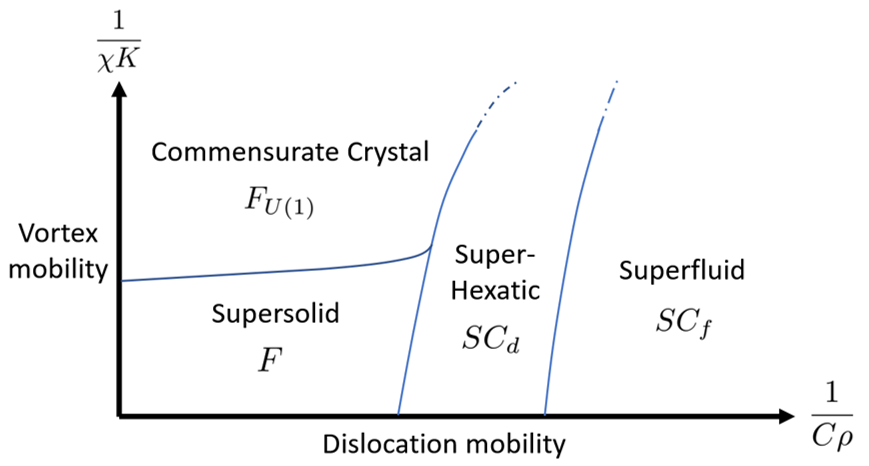}
 \caption{A schematic phase diagram illustrating phases derived from the supersolid (a $U(1)$-symmetry broken fracton phase, $F$).  Upon condensation of appropriate defects, bosons can transition to a commensurate crystal (a $U(1)$-symmetric fracton phase, $F_{U(1)}$), super-hexatic, or superfluid phase.  Note that $U(1)$-symmetric liquid and hexatic phases are forbidden at zero temperature.}
 \label{fig:phase}
\end{figure}

In this Letter, we address these basic issues by starting with a model of a two-dimensional supersolid ($i.e.$ an ``incommensurate'' crystal), characterized by intertwined crystalline and off-diagonal U(1) orders, driven by condensation of vacancy/interstitial defects in the crystal's ground state.  We perform a duality transformation on this model to arrive at a hybrid vector-tensor $U(1)$ fracton gauge theory, combining aspects of both fracton-elasticity and particle-vortex duality. This gauge dual exhibits ``mutual'' axion electrodynamics between the vector and tensor sectors of the theory, along with an associated generalized Witten effect.\cite{witten,wilczek}  The broken particle-hole symmetry of the supersolid, restored only in the commensurate crystal, corresponds to an effective ``time-reversal'' symmetry breaking of the gauge dual, akin to the role of particle-hole symmetry in Son's theory of a half-filled Landau level.\cite{son,chong,chong2}

The vector-tensor gauge dual of the supersolid describes a fracton phase in which dipoles are fully mobile, endowed by the broken $U(1)$ symmetry of atom number conservation.  The supersolid then serves as a ``mother" state from which other phases can be obtained via defect condensation transitions.  Vortex condensation drives a quantum transition from the supersolid to the commensurate crystal, which restores $U(1)$ symmetry and confines dipoles to be one-dimensional, corresponding to the familiar dislocation glide-only constraint.\cite{landau,chaikin}  Meanwhile, condensations of fracton dipoles and charges, respectively, lead to gauge dual descriptions of super-hexatic and superfluid phases.  Because dislocations and disclinations carry $U(1)$ boson number (as encoded in the generalized Witten effect, to be discussed below), proliferation of these defects at quantum melting transitions automatically condenses vacancies/interstitials, leading to off-diagonal superfluid order in the quantum hexatic and isotropic liquids.  Furthermore, this charge attachment allows a direct transition between the commensurate crystal and super-hexatic phases, characterized by distinct Landau order parameters.  This transition, if direct and continuous, lies outside the Landau-Ginzburg framework, drawing an intriguing connection with the physics of deconfined quantum criticality.\cite{dqcp,defined}

\emph{Duality Transformation.}  To faithfully capture the low-energy degrees of freedom of a supersolid, we represent the bosonic atom field annihilation operator, $\hat\psi(\xv) = \hat\psi_0 + \sum_{\G} \hat\psi_{\G}e^{i{\G\cdot\xv}}$ in terms of long wavelength and reciprocal lattice ($\G$) components $\hat\psi_0 = \sqrt{\hat n_0} e^{i\hat\phi}$ and $\hat\psi_\G = \sqrt{\hat n_\G} e^{i\hat\phi+i\G\cdot\hat{\bf u}}$, respectively. The density operators $\hat n_0$, $\hat n_\G \equiv n_\G^0 + \G^{-1}\cdot\hat{\vpi}$ and the corresponding canonically conjugate phases $\hat\phi$, $\G^{-1}\cdot\hat{\bf u}$ describe the low-energy (superfluid phase and phonon) excitations of the superfluid and crystalline orders, respectively.  The associated Hamiltonian density, consisting of the kinetic and interaction components, is given by:
\begin{eqnarray}
\hat{\cal H} &=& \frac{1}{2}\rho^{-1}\hat\pi^2
+\frac{1}{2}\tilde C^{ijk\ell}\hat u_{ij}\hat u_{k\ell}
+\frac{1}{2}\tilde K(\grad\hat\phi)^2
+\frac{1}{2}\chi^{-1}\hat n^2\nonumber\\
&& -\mu\hat n + \tilde{g}_1 \grad\hat\phi\cdot\hat\vpi + \tilde{g}_2 \hat n \hat u_{ii}\ ,
\end{eqnarray}
where $\hat\vpi$ and $\hat n=\hat n_0 + \sum_\G\hat n_\G$ are the momentum and number density, $\hat u_{ij} = \frac{1}{2}(\partial_i\hat u_j + \partial_j\hat u_i)$ is the symmetrized linear strain tensor, $\mu$ the chemical potential, $\rho$ the boson average mass density, $\tilde K$ the superfluid stiffness, $\chi$ the compressibility, and $\tilde C^{ijk\ell}$ the tensor of elastic coefficients\cite{chaikin,landau}, capturing elasticity and dynamics of the quantum crystal. As usual, the appearance of the antisymmetric strain, $\epsilon^{ij}\partial_iu_j$ in the Hamiltonian is forbidden by the underlying rotational invariance.  The $\tilde{g}$-terms are the symmetry-allowed current- and number-density couplings between the elastic and superfluid components.  At the microscopic level, the $n$-$\phi$ sector encodes condensation of vacancy/interstitial defects that allows for superfluidity to coexist with crystalline order, breaking respective global $U(1)$ and spatial symmetries.

The dual description of the physics is most efficiently derived in an equivalent coherent-state path-integral formulation, $Z = \int [d\vpi][d{\bf u}][d n][d\phi] e^{i S}$, with the action $S =\int_{\xv,t}\left[\vpi\cdot\partial_t{\bf u} - n\partial_t\phi -  \cH[\vpi,{\bf u}, n, \phi]\right]$, (with $\int_{\xv,t}\equiv\int d^2x dt$, $\hbar = 1$),
\begin{eqnarray}
S =\int_{\xv,t}\bigg[\oh\rho(\partial_t{\bf u})^2 
- \oh C^{ijk\ell} u_{ij} u_{k\ell}
+\oh\chi (\partial_t\varphi)^2 \nonumber \\
- \oh K (\grad\varphi)^2 - g_1\partial_t{\bf u}\cdot\grad\varphi
+ g_2\partial_t\varphi\grad\cdot{\bf u}\bigg]\ .
\label{St}
\end{eqnarray}
Above we defined a shifted phase field $\varphi = \phi - \mu t$, stiffnesses $K = \tilde K - \rho \tilde{g}_1^2$ and $C_{ijk\ell}=\tilde C_{ijk\ell} - \chi \tilde{g}_2^2\delta_{ij}\delta_{k\ell}$, and couplings $g_1 = \tilde{g}_1\rho$ and $g_2 = \tilde{g}_2\chi$.  To develop physical intuition for the cross terms and constrain the coefficients $g_{1,2}$, we analyze the equation of motion for $\varphi$,
\begin{eqnarray}
\partial_t n_d + \grad\cdot{\bf j}_d = 
g_2\partial_t\grad\cdot{\bf u}-g_1\grad\cdot\partial_t{\bf
  u}\equiv J_s,
\label{ndBreakdown}
\end{eqnarray}
where we have identified the net vacancy/interstitial defect number $n_d = n_i - n_v$ and current density ${\bf j}_d$ as
\begin{eqnarray}
n_d &=& -\chi\partial_t\varphi = n + g_2\grad\cdot{\bf u},\\
{\bf j}_d &=& K \grad\varphi = {\bf j} - g_1\partial_t{\bf u},
\end{eqnarray}
In terms of total atom number $n$ and current density ${\bf j}$, the equation of motion simply reduces to the full boson continuity equation $\partial_t n + \grad\cdot{\bf j} = 0$.  The source term $J_s$ represents the non-conservation of vacancies/interstitials.  In the absense of topological lattice defects ($i.e.$ for single-valued lattice distortions, such that $\partial_i\partial_t{\bf u}=\partial_t\partial_i{\bf u}$), the number difference $n_d$ is conserved, which thereby requires $J_s=0$, corresponding to $g_1 = g_2\equiv g$.  In the presense of dislocations (nonzero Burgers vector density $\grad\times\grad u_i = b_i$), $J_s$ is generically nonzero and is given by the trace of the dislocation current tensor $J_s^{ij} = \epsilon^{ik}\epsilon^{j\ell}(\partial_t\partial_k - \partial_k\partial_t)u_\ell = g(\zh\times{\bf v})_{(i} b_{j)}$.\cite{interstitials,elasticity}  Eq.\rf{ndBreakdown} then dictates that dislocation climb (motion transverse to its Burgers vector ${\bf b}$) leads to vacancy/interstitial creation/annihilation, as dictated by microscopics.\cite{interstitials,elasticity}

It is also enlightening to examine the phonon equation of motion,
\begin{eqnarray}
\hspace{-1cm}
\rho\partial_t^2u_i - C^{ijk\ell}\partial_ju_{k\ell} 
&=& g(\partial_t\partial_i 
- \partial_i\partial_t)\varphi \equiv g\epsilon_{ij} j^j_v\ ,
\label{newton}
\end{eqnarray}
where ${\bf j}_v = {\bf v} n_v$ is the current of vortices moving with velocity $\bf v$ in the vacancy/interstitial condensate. This Newton's equation reveals that, while total momentum ${\vpi} = \rho\partial_t{\bf u} - g \grad\varphi$ (with the total stress $C^{ijk\ell}u_{k\ell}+g n_d\delta_{ij}$) is conserved, the phonon momentum can change due to vortex motion, which causes decay of supercurrents and imparts a force $g\zh\times{\bf j}_v$ on the lattice.

To obtain a dual gauge theory description of the effective action, $S$, for
a supersolid, it is convenient to introduce Hubbard-Stratonovich fields $n$,
$\pi_i$, $\sigma_{ij}$, and $j_i$, with the resulting action,
\begin{align}
&S =\int_{\xv,t}\bigg[\pi^i \dot{u}_i - \oh\bar\rho^{-1}\pi^2
+\oh\bar C^{-1}_{ijk\ell}\sigma^{ij}\sigma_{k\ell} -
\sigma_{ij}u_{ij} - n\dot{\varphi} \nonumber\\
& - \oh\bar\chi^{-1} n^2
+\oh\bar K^{-1} j^2 - j^i\partial_i\varphi - \bar g\pi^i j_i - \underline{g}C^{-1}_{iik\ell} \sigma^{k\ell} n\bigg]\ ,
\label{S_hs}
\end{align}
where dots denote time derivatives, and the ``barred'' couplings are shifted counterparts of those appearing in Eq. \rf{St}, such that the original action is recovered when all four Hubbard-Stratonovich fields are integrated out.  To execute the duality and obtain a physical interpretation of these dual fields, we break up the phonon $u_i=\tilde u_i + u_i^{(s)}$ and superfluid phase $\varphi=\tilde \varphi + \varphi^{(s)}$ into a single-valued ($\tilde u_i, \tilde \varphi$) and singular ($u_i^{(s)}, \varphi^{(s)}$) components, respectively. Because $\tilde u_i, \tilde\varphi$ enter the action linearly, they can be integrated out of the partition function, leading to local constraints on total momentum (Newton's law) and on the total atom number conservation, giving the continuity equations $\partial_t \pi^i - \partial_j\sigma^{ij} = 0$ and $\partial_t n + \grad\cdot{\bf j} = 0$, consistent with the equations of motion found earlier in Eqs. \rf{ndBreakdown} and \rf{newton}.  While lattice momentum and vacancy/interstitial number are not independently conserved, they are exactly conserved for the {\em total} atom momentum and the {\em total} number of atoms (combined vacancy/interstitial and lattice). The corresponding total currents are the stress $\sigma^{ij}$ and total atomic current ${\bf j}$.

To solve these continuity constraints, we introduce rotated field redefinitions: $\pi^i = \epsilon^{ij}B_j$, $\sigma^{ij} = -\epsilon^{ik}\epsilon^{j\ell}E_\sigma^{k\ell}$, $n = b$, and $j^i = \epsilon^{ij}e_j$.  In terms of these fields, the constraints take the form of Faraday equations:
\begin{equation}
\partial_t B^i + \epsilon_{jk}\partial^jE_\sigma^{ki} = 0\,\,\,,\,\,\,\,\,\,\,\,\,\,\,\,\,\,\,\,\,\,\partial_t b + \epsilon_{jk}\partial^j e^k = 0
\end{equation}
which (as in standard electrodynamics and in basic boson-vortex duality) can be solved by symmetric tensor and vector gauge potentials, in terms of which the fields take the form $B^i = \epsilon_{jk}\partial^jA^{ki}$, $E^{ij}_\sigma = -\partial_t A^{ij} - \partial_i\partial_j A_0$, $b = \epsilon^{ij}\partial_ia_j$, and $e^i = -\partial_t a^i - \partial^i a_0$.  Plugging these forms back into Eq. \rf{S_hs}, we obtain the dualized action, taking the form of a hybrid vector-tensor gauge theory,
\begin{align}
&S =\int_{\xv,t}\bigg[\oh\hat C_{ijk\ell} E^{ij} E^{k\ell} -
\oh\bar\rho^{-1} B^2+\oh\bar K^{-1}e^2 - \frac{1}{2}\bar\chi^{-1} b^2\nonumber\\
&- \bar g {\bf B}\cdot{\bf e} - \underline{g} E^{ii} b - J_s^{ij} A_{ij} - n_s A_0 - {\bf j}_v\cdot{\bf a} - n_v a_0\bigg]\ ,\nonumber\\
\label{S_dual}
\end{align}
where $\hat C_{i j k \ell}= \epsilon^{is}\epsilon^{jt}\epsilon^{kn}\epsilon^{\ell m} \bar C_{s t n m}$, $E_{ij} = \hat{C}^{-1}_{ijk\ell}E_\sigma^{k\ell}$ is the canonical conjugate to $A_{ij}$, $J_s^{ij}$ is the tensor dislocation current described earlier, and $n_s = \epsilon^{ik}\epsilon^{j\ell}\partial_i\partial_ju_{k \ell}^{(s)} = s + \zh\cdot\grad\times{\bf b}$ is the total disclination density, sourced by elementary disclinations $s =
\epsilon_{ij}\partial_i\partial_j\theta^{(s)}$ ($i.e.$ windings of the bond angle $\theta =\oh\epsilon_{ij}\partial_iu_j^{(s)}$) and the curl of the dislocation density, ${\bf b}=\epsilon_{ij}\partial_i\partial_j{\bf u}^{(s)}$.  The tensor sector takes the form of the scalar charge fracton tensor gauge theory, with disclinations $n_s$ acting as fracton charges, obeying conservation of both charge and dipole moment.  Similarly, $n_v$ is the vortex number, defined by $n_v =\epsilon^{ij}\partial_i\partial_j\varphi^{(s)}$, and the vortex current density ${\bf j}_v$ is defined in Eq. \rf{newton}.

We observe that on this dual gauge theory side, the boson-elastic cross-coupling of the quantum crystal manifests itself as generalized ``axion'' terms featuring products of electric and magnetic fields.\cite{wilczek}  Note that these axion terms are allowed in two spatial dimensions, instead of the usual three, due to the presence of extra tensor indices.  In close analogy with the Witten effect\cite{witten}, we expect such axion terms to lead to charge attachment, manifesting in modified Gauss's laws in the theory.  Indeed, by varying the action with respect to $a_0$, we obtain the Gauss's law for dual electric charges (vortices) as:
\begin{equation}
\grad\cdot{\bf e} = n_v - \bar g\grad\cdot{\bf B},
\end{equation}
corresponding to attachment of magnetic flux of the dual tensor gauge theory to the dual electric charges (vortices). In the supersolid language, we have $\grad\cdot{\bf B} =\grad\times\vpi$, indicating that the above flux attachment corresponds to the crystal's angular momentum contribution to the vacancies/interstitials' vorticity.

Similarly, the tensor Gauss's law for fractons
\begin{equation}
\partial_i\partial_j E^{ij} = n_s + \underline{g} \hat C^{-1}_{i i k
  \ell}\partial_k\partial_\ell b
\label{discgauss}
\end{equation}
is modified by an inhomogeneous magnetic flux density.  Since the latter corresponds to boson density, this modified Gauss's law encodes the fact that lattice defects carry boson number.  In more detail, the second derivative structure of the right-hand side indicates that bosons are attached to quadrupoles of disclinations (fractons).  As we elaborate below, this feature is crucial in excluding unphysical Mott-insulating phases of bosons when crystal defects have melted the lattice.

\begin{figure}[t!]
 \centering
 \includegraphics[scale=0.3]{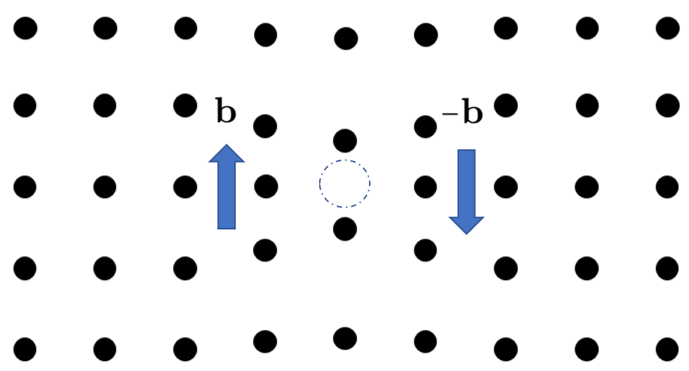}
 \caption{A disclination quadrupole, constructed as a bound state of two equal and opposite dislocations with Burgers vectors $\bf b$ and $-\bf b$, carries a unit of vacancy number, as can be seen by the deficiency of a single atom in the middle of the configuration.}
 \label{fig:vacancy}
\end{figure}

While the charge attachment effects are similar, these generalized axion terms have several notable differences from the conventional axion term of Maxwell theory, such as the absence of topological invariance or periodicity.  Since these concepts play no role in the physical theory of supersolids, we relegate discussion of most of these differences to the Supplemental Material.  For now, we note the unusual symmetry properties of these terms, which are even under both time reversal and spatial inversion ($e^i\rightarrow -e^i$, $B^i\rightarrow -B^i$, $b$ and $E^{ij}$ unchanged).  Instead, they violate particle-hole symmetry ($b\rightarrow -b$, $B^i\rightarrow -B^i$, $e^i$ and $E^{ij}$ unchanged), which acts as an effective ``time-reversal," as encountered previously in discussions of the half-filled Landau level.\cite{son,chong,chong2}  This is consistent with the fact that the particle-hole symmetry of a commensurate crystal is generically spontaneously broken in a supersolid via condensation of vacancies or interstitials.  As such, there is no symmetry consideration which forces $g$ to take a particular quantized value.  Instead, $g$ takes a nonuniversal value which is determined by the microscopic condition that a bound state of two equal and opposite dislocations separated by a single lattice site carries one unit of vacancy number (Fig.\ref{fig:vacancy}).

\emph{Fracton phases and quantum transitions.} In the previous section, we established a dual vector-tensor $U(1)$ gauge theory of a supersolid, with the two sectors coupled by axion-like terms.  We can explicitly account for the quantum dynamics of the dual matter fields (vortices of the condensate, and disclinations and dislocations of the quantum crystal) in terms of a coupled vector-tensor dual ``superconductor," with the action $S= S_{Max} + S_c$, where $S_{Max}$ is the Maxwell piece of the action for both vector and tensor sectors, and:
\begin{align}
S_c& = \int_{\xv,t}\bigg[\mu b - \overline{g}{\bf B}\cdot{\bf e} - \underline{g} E_{ii} b -\cos(\dot{\tilde\phi} - a_0) - \cos(\dot{\tilde\theta} - A_0)   \nonumber\\
&+ \eta\cos(\grad\tilde\phi- {\bf a}) +\eta'\sum_n\cos(\alpha^{ij}_{(n)}(\partial_i\partial_j\tilde\theta - A_{ij}))\bigg],
\label{Ssc_dual}
\end{align}
where dots denote time derivatives and $\alpha_{ij}^{(n)} = a^2b^{(n)}_ib^{(n)}_j$ encodes the glide-only constraint of dislocations (with lattice constant $a$).  The model exhibits a variety of quantum phases with distinct patterns of dual matter (topological defect) condensation.  The Coulomb phase of the tensor gauge sector (corresponding to crystalline order, $i.e.$, a vacuum of disclinations and dislocations) displays two qualitatively distinct fractonic quantum states.

\begin{figure}[t!]
 \centering
 \includegraphics[scale=0.33]{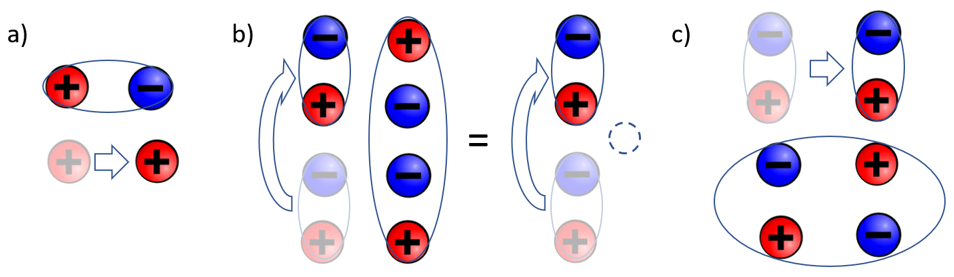}
 \caption{a) Fracton motion is forbidden as it requires emission of a conserved dipole.  b) \emph{Longitudinal} dipole motion (dislocation climb) is forbidden as it requires emission of a linear quadrupole carrying conserved vacancy/interstitial number, corresponding to local compression of the crystal, $i.e.$, a vacancy defect.  c) \emph{Transverse} dipole motion (dislocation glide) is allowed as it creates a square quadrupole, corresponding to local shear.}
 \label{fig:quad}
\end{figure}

(i) One Coulomb phase of the tensor gauge theory is a {\em U(1) symmetric fracton state}, $F_{U(1)}$, that is in the Higgs phase of the vector gauge theory, corresponding to condensed vortices, describing a commensurate Mott-insulating crystal.  In this phase, the gapped vector gauge field ${\bf a}$ can be safely integrated out, leading to a gapless tensor-only gauge theory in its Coulomb phase, previously derived in Ref. \onlinecite{elasticity}. It exhibits immobile fracton charges (disclinations) and subdimensional fracton dipoles, $\bf p = {\bf \hat{z}}\times\bf b$ (dislocations).  These fractonic mobility restrictions can be understood in terms of conservation laws, which include conservation of fracton charge ($\int_{\bf x} \,n_s = \textrm{constant}$) and dipole moment ($\int_{\bf x} \,n_s\bf x = \textrm{constant}$), as well as enrichment by the global $U(1)$ symmetry of atom number conservation.  The latter is encoded in the relation between diagonal components of the fracton quadrupole $E_{ij}$ and the vacancy/interstitial density:
\begin{equation}
\int_{\bf x} n_s x^2 - 2n_d = \textrm{constant}.
\end{equation}
The dipoles can freely move transversely to $\bf p$ (dislocation glide), which corresponds to creation of square fracton quadrupoles $E_{xy}$ (see Fig. \ref{fig:quad}), that are not conserved. However, a dipole's motion along $\bf p$ (dislocation climb, transverse to the Burgers vector $\bf b$) generates linear quadrupoles, $E_{ii}$, that corresponds to creation of vacancies/interstitials, and is thus forbidden by the global $U(1)$ symmetry enforcing atom number conservation. Thus, the subdimensional character of dipoles inside the $F_{U(1)}$ fracton state is $U(1)$ symmetry-protected.  This provides a concrete example of a symmetry-enriched fracton phase, in which the mobility constraints become more restrictive in the presence of a global symmetry.

(ii) The other Coulomb phase of the tensor gauge theory is a {\em U(1) symmetry-broken fracton state}, $F$, that is in the Coulomb phase of the vector gauge theory, corresponding to gapped vortices (condensed vacancies/interstitials), describing a supersolid.  This supersolid fracton Coulomb phase is characterized by coupled gapless vector, $\bf a$, and tensor, $A_{ij}$, gauge fields, with the action given in Eq.\rf{S_dual}.  It exhibits immobile fracton charges (disclinations), but unlike $F_{U(1)}$, inside $F$ the breaking of $U(1)$ symmetry liberates dipoles, $\bf p$ (dislocations) to be fully mobile.  The $F_{U(1)}$-$F$ transition between these two fracton phases is thus driven by global $U(1)$ symmetry breaking or equivalently by a Coulomb-to-Higgs phase transition in the associated vector gauge theory sector.

In addition to the $F_{U(1)}$ and $F$ fracton phases -- the Coulomb (crystalline) phases of the tensor gauge theory sector -- the model \rf{Ssc_dual} also admits several non-fractonic states associated with the complementary Higgs phase of the tensor gauge theory sector.  These are driven by condensation of dual matter such as dipoles (dislocations) and fractons (disclinations), corresponding to various ``superconductor" phases of the tensor gauge theory.  The dipole superconductor, $SC_d$, is dual to a super-hexatic phase, while a fracton superconductor, $SC_f$, corresponds to the ordinary dual description of a superfluid.\cite{ber1,ber2,KT,halperin,nelson,young,chaikin,landau}  The presence of condensates in these tensor superconductors mobilizes the charges, precluding ``fracton order."  A schematic phase diagram is depicted in Figure \ref{fig:phase}.

Furthermore, as illustrated in Fig.\rf{fig:vacancy}, a pair of equal and opposite dislocations a lattice constant apart, is equivalent to an atom vacancy (or, in reverse configuration, an interstitial) and thereby carries an atom number ``charge." Formally, the Hamiltonian thus admits an operator $\hat b^\dagger_{\bf b}\hat b^\dagger_{-\bf b}\hat a + h.c.$, where two dislocations annihilate into a vacancy and vice versa. A condensation of $b$'s therefore clearly requires atomic off-diagonal-long-range order.  Thus, at zero temperature these non-fractonic hexatic and isotropic fluid phases of condensed dislocations and disclinations are guaranteed to be superfluid. The lack of crystalline order precludes commensuration necessary for a Mott-insulator phase of vacancies/interstitials and for the associated vortex condensation.  On the dual side, this guarantees that the non-fractonic Higgs phases of the tensor gauge theory sector necessarily break $U(1)$ symmetry, and must be in the Coulomb phase of the vector gauge theory.  This interplay between crystalline and $U(1)$ symmetries is the same as that observed in the context of deconfined quantum critical points between phases with different symmetry breaking (such as the N\'eel-VBS transition), in which defects of one ordered state carry the quantum number of the other order parameter, allowing for a direct phase transition outside the Landau-Ginzburg framework.\cite{dqcp,defined}  Similar charge attachment to lattice defects has also been observed in the context of melting transitions in quantum Hall phases.\cite{hall}

\emph{Conclusions.}  In this work, we have derived a dual hybrid vector-tensor gauge theory formulation of a quantum crystal, that accounts for dynamics of atomic vacancies and interstitials.  This duality combines the newly established fracton-elasticity duality\cite{elasticity} with conventional particle-vortex duality\cite{stone,dasgupta,fisher} into one ``mother'' duality, featuring both vector and tensor gauge fields, coupled by ``mutual'' axion terms and exhibiting an associated generalized Witten effect.  We thereby establish the existence of two qualitatively distinct fracton phases, corresponding to a commensurate and incommensurate (supersolid) crystals, distinguished by $U(1)$ symmetric and broken states of atom number symmetry, with the phases exhibiting fracton charges and, respectively, subdimensional and fully mobile dipoles.  A condensation of fracton dipoles provides a gauge dual description of phase transition into a superhexatic, and vortex condensation restores $U(1)$ symmetry, confining dipoles to be subdimensional, and thereby drives a transition between $F$ and $F_{U(1)}$ fracton phases.  We leave further exploration of these and related phases to future studies.

\emph{Note:}  As this manuscript was being completed, we became aware of a related paper by A. Kumar and A. C. Potter, which performs complementary work on a similar topic.\cite{kumar}

\vspace{0.2cm}

\begin{acknowledgments}
We gratefully acknowledge useful conversations with Mike Hermele, Matthew Fisher, Andrew Potter, Ajesh Kumar, T. Senthil, Shriya Pai, Brian Swingle, Rahul Nandkishore, Yizhi You, Abhinav Prem, Yang-Zhi Chou, and Han Ma.  This work was supported by a Simons investigator award to Leo Radzihovsky.  MP is also supported by the NSF grant 1734006 and by the Foundational Questions Institute (fqxi.org; grant no. FQXi-RFP-1617) through their fund at the Silicon Valley Community Foundation.
\end{acknowledgments}

\end{document}


\title{Supplemental Material for ``Symmetry Enriched Fracton Phases from Supersolid Duality"}
\author{Michael Pretko and Leo Radzihovsky \\
\emph{Department of Physics and Center for Theory of Quantum Matter} \\
\emph{University of Colorado, Boulder, CO 80309}}
\date{\today}
\maketitle

\section*{Appendix: Properties of the Generalized Axion Terms}

In the main text, we showed that a supersolid has a gauge dual description given by the following hybrid vector-tensor gauge theory:
\begin{equation}
S = S_{Max} + S_{source} + S_g
\end{equation}
where:
\begin{equation}
S_{Max} = \int_{x,t}\bigg[\frac{1}{2}\hat C_{ijk\ell} E^{ij} E^{k\ell} -
\frac{1}{2}\bar\rho^{-1} B^2+\frac{1}{2}\bar K^{-1}e^2 - \frac{1}{2}\bar\chi^{-1} b^2\bigg]
\end{equation}
is the Maxwell sector of the two gauge theories and
\begin{equation}
S_{source} = \int_{x,t} \bigg[- J_s^{ij} A_{ij} - n_s A_0 - {\bf j}_v\cdot{\bf a} - n_v a_0\bigg]
\end{equation}
provides the source terms.  Such terms are to be expected.  The more intriguing aspect of this dual gauge theory is the presence of generalized axion terms, given by:
\begin{align}
S_g =\int_{x,t}\bigg[ - \bar g {\bf B}\cdot{\bf e} - \underline{g} E^{ii} b \bigg]
\end{align}
which creates nontrivial interplay between the electric and magnetic sectors of the theory.  This portion of the action has a strong similarity the conventional $\vec{E}\cdot \vec{B}$ axion terms of ordinary $U(1)$ gauge theories, and indeed they share some properties.  However, in this Appendix, we discuss some of the notable differences between the present action and conventional axion physics.

The most important similarity between the generalized axion terms and more conventional ones lies in the physics of charge attachment.  As discussed in the main text, the primary physical effect of the axion piece of the action is to create a generalized Witten effect by modifying the Gauss's laws of the theory:
\begin{equation}
\partial_ie^i = n_v - \bar g\partial_iB^i
\end{equation}
\begin{equation}
\partial_i\partial_j E^{ij} = n_s + \underline{g} \hat C^{-1}_{i i k
  \ell}\partial_k\partial_\ell b
\end{equation}
In supersolid language, the first equation indicates that the vortices of the vacancy/interstitial condensate carry lattice angular momentum of the elastic sector.  Meanwhile, the second equation tells us that the topological lattice defects ($i.e.$ disclinations and dislocations) carry nontrivial vacancy number of the underlying atoms.  This charge attachment has important consequences for the physics of quantum phases of bosons, such as ruling out trivial gapped phases, in accordance with the Lieb-Schultz-Mattis theorem.

While the corresponding Witten effects are similar, there are also many notable differences between the generalized axion terms and more conventional ones.  Importantly, the action of symmetries on this vector-tensor gauge theory are somewhat unusual.  Under time reversal, the physical fields transform as follows:
\begin{align}
e^i&\rightarrow -e^i\,\,\,\,\,\,\,\,\,\,\,\,\,\,\,B^i\rightarrow -  B^i\nonumber \\
&E^{ij}\rightarrow E^{ij}\,\,\,\,\,\,\,\,\,\,\,\,\,\,\,b\rightarrow b
\end{align}
As such, the axion terms are even under time reversal, which therefore does not constrain the value of $g$.  This is in contrast with the conventional axion $\theta$ parameter, for which time reversal symmetry dictates that $\theta = 0$ or $\pi$, since $\theta \rightarrow -\theta$ under time reversal and $\theta$ is a $2\pi$-periodic variable.

While symmetric under time reversal, the axion action fails to be invariant under particle-hole symmetry, which has the following action:
\begin{align}
e^i\rightarrow e^i\,\,\,\,\,\,\,\,\,\,\,\,\,\,\,B^i\rightarrow -B^i\nonumber \\
E^{ij}\rightarrow E^{ij}\,\,\,\,\,\,\,\,\,\,\,\,\,\,\,b\rightarrow -b
\end{align}
We therefore see that particle-hole symmetry acts as an effective time reversal, as seen in earlier treatments of boson-vortex duality.  The failure of the axion terms to be invariant under this symmetry reflects the fact that the particle-hole symmetry of a commensurate crystal is spontaneously broken in a supersolid.  Since the supersolid is not particle-hole symmetric, there is once again no constraint on the value of $g$.  As discussed in the main text, the value of $g$ is determined by the microscopics by the fact that the minimal quadrupole ($i.e.$ two dislocations separated by one lattice spacing) should carry a single unit of vacancy charge.  Of course, this notion relies on having a smallest fundamental Burgers vector in the theory, which amounts to dipole moment being quantized.  Such quantization arises due to the compactness of the underlying microscopic variables.  For example, if atoms are shifted by a Bravais lattice vector, one can recover the original lattice of the system (appropriately accounting for the change in boson number at each site).  This further demonstrates the point that the value of $g$ is determined by non-universal microscopic data, as opposed to being quantized to some universal value.

Another important feature of the generalized axion terms is absence of periodicity of the action with respect to $g$.  For an ordinary axion term, characterized by its $\theta$ angle, the action is invariant under $\theta\rightarrow\theta + 2\pi$.  In the supersolid, however, there is no such invariance.  For an ordinary axion term, this periodicity can be understood directly in the language of charge attachment.  When $\theta$ is a multiple of $2\pi$, an integer amount of electric charge has been attached to a magnetic monopole.  At this point, one can simply bind an appropriate number of electrons (or holes) to the monopole to undo the charge attachment and recover the original system.  Crucially, however, this relies on the existence of having ``bare" charges in the system in addition to the dyons (electric-magnetic bound states) governed by the Witten effect.  In a system made up of ordinary electrons, bare electrons are always part of the Hilbert space.  In the present context, however, there is no such thing as a ``bare" vacancy or ``bare" topological lattice defects.  Rather these objects are fundamentally tied together.  As such, there is no binding procedure which can undo the charge attachment effected by the axion term.  Therefore, the system is not periodic as a function of the parameter $g$.

Finally, we note that there is no reason to expect that the generalized axion terms are topologically invariant.  Ordinary axion terms are completely insensitive to the metric of spacetime, just like the action for topological phases of matter.  Fracton phases, however, are more sensitive to geometry than more conventional systems.  For example, gapped fracton systems can have extra ground state degeneracy induced by curvature \cite{field}, and the higher rank tensor gauge theories may have their higher moment conservation laws weakly violated on curved spaces \cite{curved}.  The present model is no exception.  In contracting indices, we have always implicitly made use of the flat space metric, and we do not have any of the special properties of ordinary axion terms (such as a relationship with Chern-Simons theory) which would give a path to metric-independence.  There is therefore no indication of any topological invariance associated with the generalized axion terms.